\documentclass[manuscript]{aastex}

\shorttitle{flare}
\shortauthors{Takasao \& Shibata}

\usepackage{bm}
\usepackage{amsmath}
\usepackage{url}

\begin{document}

%% LaTeX will automatically break titles if they run longer than
%% one line. However, you may use \\ to force a line break if
%% you desire.

\title{ABOVE-THE-LOOP-TOP OSCILLATION AND QUASI-PERIODIC CORONAL WAVE GENERATION IN SOLAR FLARES}

%% Use \author, \affil, and the \and command to format
%% author and affiliation information.
%% Note that \email has replaced the old \authoremail command
%% from AASTeX v4.0. You can use \email to mark an email address
%% anywhere in the paper, not just in the front matter.
%% As in the title, use \\ to force line breaks.

\author{Shinsuke Takasao\altaffilmark{1,3}}
\email{takasao@kwasan.kyoto-u.ac.jp}

\author{Kazunari Shibata\altaffilmark{1,2}}

%% Notice that each of these authors has alternate affiliations, which
%% are identified by the \altaffilmark after each name.  Specify alternate
%% affiliation information with \altaffiltext, with one command per each
%% affiliation.

\altaffiltext{1}{Kwasan and Hida Observatories, Kyoto University, 
Yamashina, Kyoto 607-8471, Japan}
\altaffiltext{2}{Unit of Synergetic Studies for Space, Kyoto University, Yamashina, Kyoto 607-8471, Japan}
\altaffiltext{3}{Department of Physics, Nagoya University, Aichi 464-8602, Japan}

%% Mark off your abstract in the ``abstract'' environment. In the manuscript
%% style, abstract will output a Received/Accepted line after the
%% title and affiliation information. No date will appear since the author
%% does not have this information. The dates will be filled in by the
%% editorial office after submission.

\begin{abstract}
Observations revealed that various kinds of oscillations are excited in solar flare regions. Quasi-periodic pulsations (QPPs) in the flare emissions are commonly observed in a wide range of wavelengths. Recent observations have found that fast-mode magnetohydrodynamic (MHD) waves are quasi-periodically emitted from some flaring sites (quasi-periodic propagating fast-mode magnetoacoustic waves; QPFs). Both of QPPs and QPFs imply a cyclic disturbance originating from the flaring sites. However, the physical mechanisms remain puzzling. By performing a set of two-dimensional MHD simulations of a solar flare, we discovered the local oscillation above the loops filled with evaporated plasma (above-the-loop-top region) and the generation of QPFs from such oscillating regions. Unlike all previous models for QPFs, our model includes essential physics for solar flares, such as magnetic reconnection, heat conduction, and chromospheric evaporation. We revealed that QPFs can be spontaneously excited by the above-the-loop-top oscillation. It was found that this oscillation is controlled by the backflow of the reconnection outflow. The new model revealed that flare loops and the above-the-loop-top region are full of shocks and waves, which is different from the previous expectations based on a standard flare model and previous simulations. In this paper, we will show the QPF generation process based on our new picture of flare loops and will briefly discuss a possible relationship between QPFs and QPPs. Our findings will change the current view of solar flares to a new view in which they are a very dynamic phenomenon with full of shocks and waves.
\end{abstract}


\section{Introduction}
Solar flares are the most energetic phenomenon in the solar system, where $10^{29}$ to $10^{32}$~erg of magnetic energy stored in the corona is rapidly released by magnetic reconnection -- reconnection of magnetic field lines -- on a timescale of hours \citep{priest1982,shibata2011}. Observations have revealed that various magnetohydrodynamic (MHD) waves are commonly associated with solar flares. Because coronal waves have the potential to tell us about the local plasma condition, which is difficult to directly observe, various techniques of MHD coronal seismology have been developed \citep{tom2008,aschwanden2011,demoortel2012}.

Various kinds of oscillations excited in solar flare regions have been investigated using emissions and imaging observations \citep{wang2003,nakariakov2009,liu2014}. Quasi-periodic pulsations (QPPs) in the flare emissions with periods ranging from fractions of seconds to several minutes are commonly observed in a wide range of wavelengths \citep{nakajima1983,nakariakov2009,simoes2015}. Recent observations by the Atmospheric Imaging Assembly (AIA; \citet{lemen2012}) on the {\it Solar Dynamics Observatory} (SDO; \citet{pesnell2012}) have found that fast-mode MHD waves are quasi-periodically emitted from some flaring sites (quasi-periodic propagating fast mode magnetoacoustic waves; QPFs) \citep{liu2011}. The high-sensitive monitoring observations by AIA enable us to study the wave properties of some events in detail \citep{liu2012,yuan2013}, although the statistical characteristics still remain unclear. The observed period ranges from a few 10~s to a few 100~s.

QPPs are also found in stellar flare emissions \citep{mathioudakis2003,balona2015}. Solar coronal seismology has been applied to stellar flares to estimate the physical parameters of unresolved stellar coronae \citep{nakariakov2004,mitrakraev2005}. Coronal seismology has the potential to provide a powerful tool for investigating the stellar magnetic activity which is difficult to explore from direct imaging observations. Thus, advancing our understanding of oscillations in flares will have a great impact on both the solar and stellar physics.

Both of QPPs and QPFs imply a cyclic disturbance originating from the flaring sites. In addition, it has been pointed out that there will be a relationship between the generation of QPFs and the time variability of the flare energy release (and therefore QPPs) \citep{liu2011,liu2012,shen2012,yuan2013}. However, the physical mechanisms of QPPs and QPFs remain puzzling.

% previous theoretical studies
Models have been developed to investigate the dynamical properties of QPFs. \citet{ofman2011} utilized a three-dimensional (3D) active region model in which periodic velocity perturbations at the photospheric level are introduced. They obtained QPFs whose properties are similar to that of observed QPFs. \citet{pascoe2013} and \citet{nistico2014} studied impulsively generated fast-mode waves in a magnetic funnel geometry, and found that their dispersive nature can lead to the formation of a quasi-periodic wave train. \citet{yang2015} showed that isotropic QPF-like waves are generated by multiple plasmoid ejections (see also \citet{yokoyama1998a}).

In our previous paper \citep{takasao2015}, we succeeded in performing an MHD simulation of a flare with a high spatial resolution. By performing a set of simulations, we discovered the local oscillation above the loops filled with evaporated plasma (above-the-loop-top region) and the generation of QPFs from such oscillating regions. Unlike all previous models for coronal waves, our model includes essential physics for solar flares, such as magnetic reconnection, heat conduction, and chromospheric evaporation. The new model revealed that flare loops and the above-the-loop-top region are full of shocks and waves. This result differs from the previous expectations based on a standard flare model and is not found in previous simulations by \citet{yokoyama1998b,yokoyama2001}. Using high spatial resolution, we, for the first time, revealed that QPFs can be spontaneously excited by the above-the-loop-top oscillation. In this paper, we present the new perspective on the generation of QPFs associated with solar flares and will briefly discuss a possible relationship between QPFs and QPPs.

\section{Numerical Model}
Our model is identical to our previous flare model \citep{takasao2015} except for the horizontal domain size. The calculated domain is $0\le x \le x_{\rm max}$ and $0\le y \le y_{\rm max}$, where $x_{\rm max}=9\times10^4$~km and $y_{\rm max}=6\times 10^4$~km, respectively. This domain is resolved with a uniform $1200\times800$ grid. The center of the initial current sheet is located at the left boundary where a reflecting boundary condition is applied. Our model includes essential physics for solar flares such as magnetic reconnection, heat conduction, and chromospheric evaporation. The model atmosphere consists of a cool dense layer (chromosphere) and a hot tenuous layer (corona). The initial magnetic field is assumed to be a force-free field. The initial gas pressure is assumed to be uniform. The initial plasma $\beta$, defined as the ratio of the gas pressure to the magnetic pressure, is a free parameter. We examined the cases of $\beta=$0.06, 0.08, 0.1, and 0.2. To allow the magnetic field to reconnect, we impose a localized resistivity. The localized resistivity is fixed in time and space to realize a fast and quasi-steady magnetic reconnection with a single X-point \citep[e.g.][]{ugai1992}. This means that we neglect the oscillations caused by plasmoids and focus on the oscillations excited by a quasi-steady reconnection outflow. The thermal conductivity is set to the a value three times smaller than the Spitzer value. The numerical scheme is based on a Harten-Lax-van Leer (HLL) scheme developed by \citet{miyoshi2005}, HLLD, which is a shock-capturing scheme. 

\section{Numerical Results}
\subsection{Evolution of Flare Loops and Emission of Coronal Waves}
\begin{figure}
\epsscale{.60}
%\plotone{qpf_demo_beta008_obs_red.eps}
%\plotone{qpf_demo_beta008_obs.eps}
\plotone{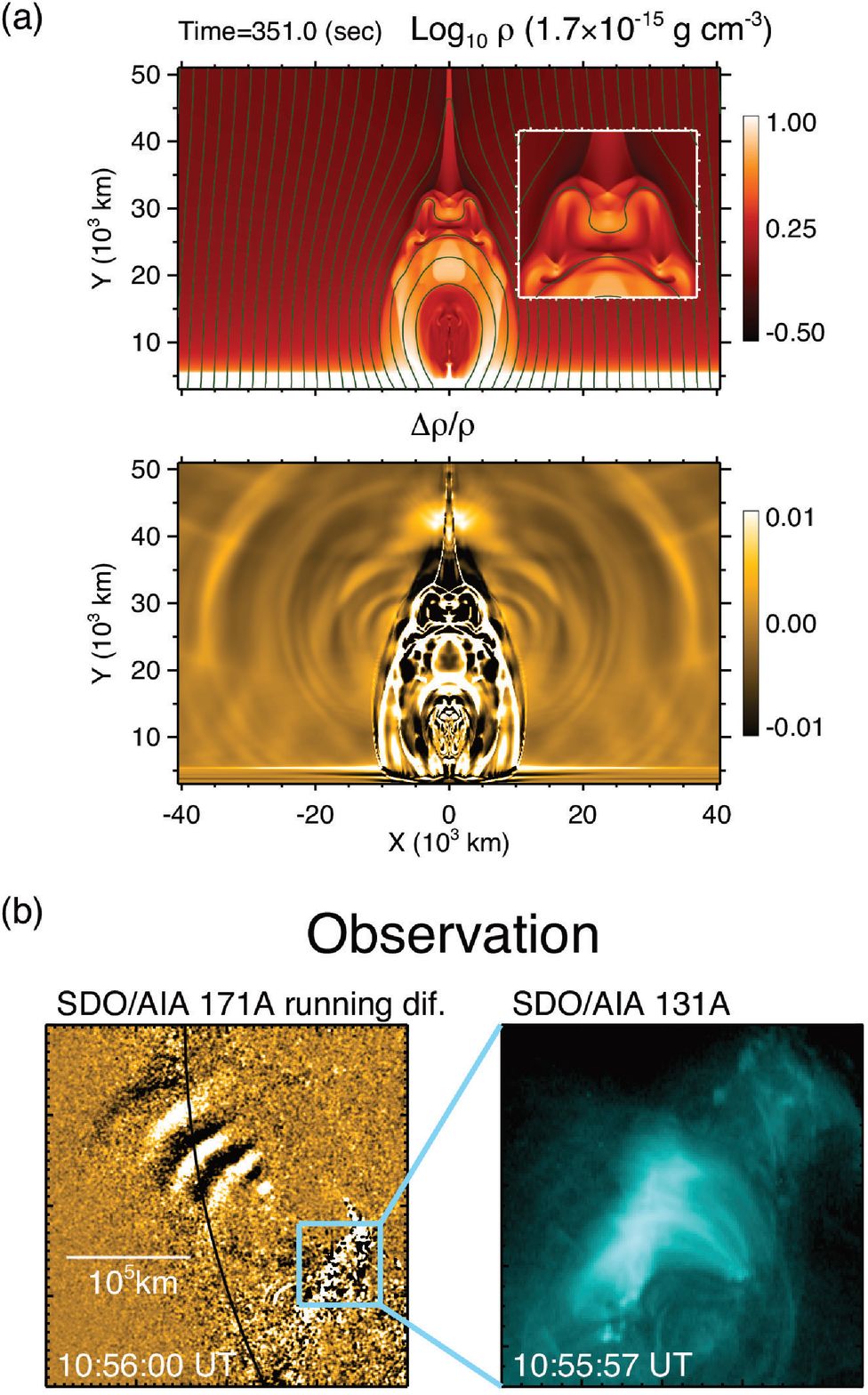}
\caption{Quasi-periodic propagating fast mode magnetoacoustic waves (QPF) in our simulation ($\beta=0.08$). Top: Density map at time=349.2 sec from the start. The solid lines denote magnetic field lines. An enlarged image of the above-the-loop-top region is also displayed. Middle: The normalized running difference of the density $\Delta \rho/\rho$, defined as $(\rho(t)-\rho(t-\Delta t))/\rho(t)$, where $\Delta t=1.8$~sec. Bottom: Observational example of QPF accompanied with a GOES C2.8 flare which occurred on 2011 30 May. An animation of the figure of the simulation is also available in the online journal.\label{fig:sim_intro}}
\end{figure}

In this paper, the case with $\beta=0.08$ is mainly mentioned as a typical example of our simulations. Figure~\ref{fig:sim_intro}(a) shows snapshots of the simulated flare. The domain where $x < 0$ is also shown, solely for visual inspection. The global evolution of the flare is essentially the same as that of our previous simulation: The reconnected magnetic field drives the Alfv\'enic outflow, and the plasma in the reconnection outflow is heated at the slow mode MHD shocks (slow shocks) emanating from the localized reconnection region \citep{petschek1964}. The reconnected fields pile up and form a loop system, which can then be filled with the hot dense plasma coming from the chromosphere (chromospheric evaporation). The loops filled with evaporated plasma will correspond to the soft X-ray flare loops. In this paper, the region above the loops filled with evaporated plasma is called ``above-the-loop-top region" (an enlarged image of this is shown in Figure~\ref{fig:sim_intro}(a)).

The normalized running difference image of the density, $\Delta \rho/\rho$, clearly shows that isotropic waves are recurrently emitted from the above-the-loop-top region (detailed analysis will be given later). The propagation speed is identical to the fast-mode MHD waves (fast waves) speed, indicating that they are fast waves. Figure~\ref{fig:sim_intro}(b) displays an observational example of QPF events accompanied by a GOES C2.8 flare which occurred on 2011 May 30 \citep[this event was studied in detail by][]{yuan2013}. The images were taken by {Atmospheric Imaging Assembly (AIA)} \citep{lemen2012} on board Solar Dynamics Observatory (SDO) \citep{pesnell2012}. The flare region showed a clear cusp-shaped loops (the 131~{\AA} channel contains the Fe~{\sc xxi} line, formed at 11~MK, which is dominant in flaring regions; it also contains lower temperature lines such as the Fe~{\sc viii} line, formed at 0.4~MK). Fast waves are quasi-periodically emitted from the flaring site during its rising phase, with a period of a few 10-100~s. These observational characteristics are similar to the simulation, although the observed anisotropic propagation of fast waves is not found in the simulations (this is probably because the coronal field structure is too simplified in our model).

\subsection{Multiple Termination Shocks and a New Picture of Flare Loops}
\begin{figure}
\epsscale{0.7}
%\plotone{oblique_horizontal_mod.eps}
\plotone{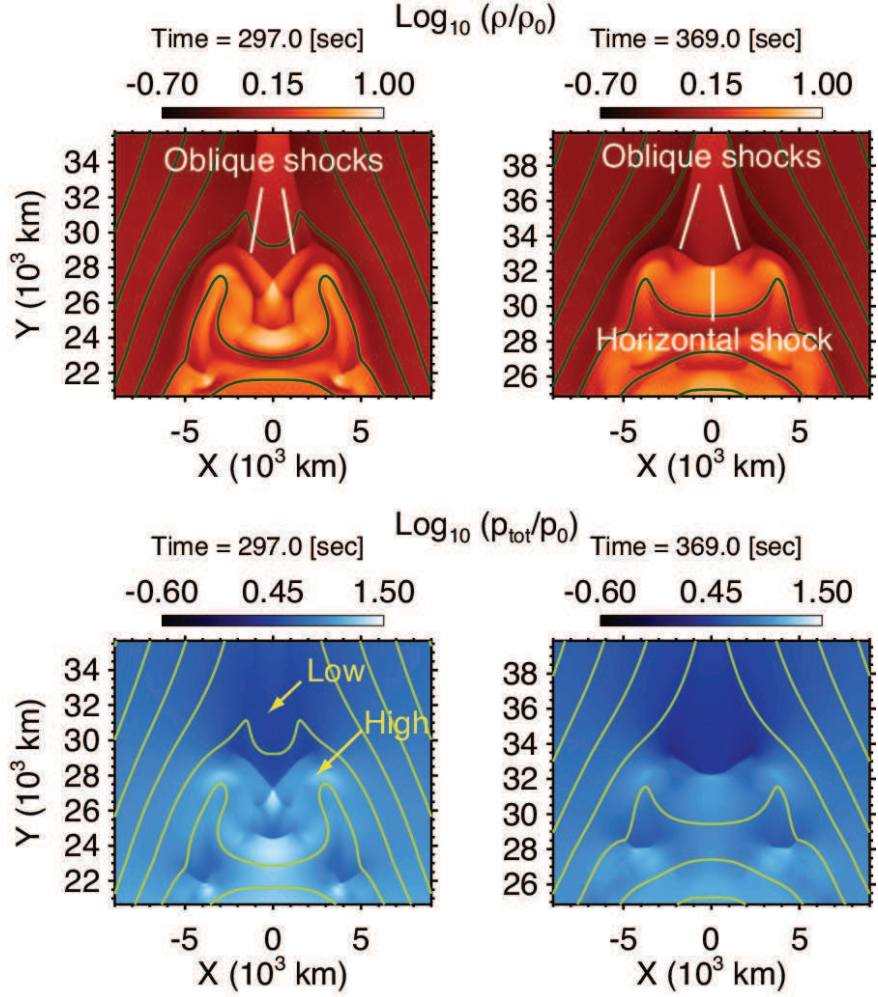}
%\plotone{test.eps}
\caption{Snapshots of the above-the-loop-top region. Top row: Density distribution. Oblique and horizontal fast shocks are indicated. Bottom row: Total pressure (gas pressure plus magnetic pressure) distribution. Note that the total pressure in the reconnection outflow is smaller than that of the above-the-loop-top region. Solid lines denote magnetic field lines. \label{fig:oblique_shock}}
\end{figure}

Since shocks in the above-the-loop-top region play an important role in the generation of coronal fast waves, we will first mention the shock structure in detail. Fast mode MHD shocks (fast shocks) are formed well above the loops filled with the evaporated plasma as a consequence of the termination of the reconnection outflow (see the top row of Figure~\ref{fig:oblique_shock}). The shocks are formed in the high-plasma $\beta$ region (see also Figure~7 of \citet{takasao2015}). In a standard flare model, a standing horizontal fast shock is expected to be formed at the termination site and is often referred to as a ``termination shock" \citep[e.g.][]{forbes1983,priest2002}. However, the simulation shows that a V-shaped pattern is formed by two oblique fast shocks and later by two oblique shocks and a single horizontal fast shock, which is very different from the standard picture. The two fast shocks sometimes reflect in the above-the-loop-top region. The ``multiple termination shocks" are also reported in our previous paper.

The two oblique shocks are formed in the following manner. The kinetic energy of the reconnection outflow is thermalized in the above-the-loop-top region. In addition, the magnetic fields are piled up there. For these reasons, the total pressure $p_{\rm tot}$ (gas pressure plus magnetic pressure) in the above-the-loop-top region is larger than the total pressure in the outflow (see the bottom row of Figure~\ref{fig:oblique_shock}). As the outflow enters the above-the-loop-top region, the ambient total pressure compresses the outflow. The external compression is caused by the two oblique fast shocks inclined at an angle to the flow. This situation is very similar to the situation often referred to as ``overexpansion" in the fluid dynamics \citep{wilson1985}, except for the existence of a magnetic field.

Heat conduction has an effect to make plasma soft: heat conduction can reduce the pressure enhanced by compression by transporting the heat elsewhere. For this reason, the push by the ambient total pressure becomes weaker in the case with heat conduction than in the case without it, leading to the formation of more vertical oblique shocks. We consider that this makes the appearance of the horizontal fast shock difficult in the cases with heat conduction. We confirmed that without heat conduction a horizontal fast shock appears at almost the same time when two oblique fast shocks are formed. The effects of heat conduction on the fast-mode Mach number of the reconnection outflow jet will be briefly discussed in the Appendix~\ref{sec:appendix}.

We tracked a specific field line  to see the history of the passage through the multiple termination shocks. The left panels of Figure~\ref{fig:te_ro_line} display snapshots of the temperature and density at a time when the shock reflection occurs. The right panels show the time-sequenced images obtained along the tracked field line. The slow and fast shocks that the field line passed are also indicated. One will find that the field line in $x>0$ always crosses more than two shocks during the period between $\sim$302~s and 315~s. It is also clear that the distance between the slow shock (attached to the reconnection outflow) and the topmost oblique fast shock becomes smaller as time progresses (indicated by arrows in the density map). Finally, we note that the temperature ahead of the topmost oblique fast shocks is enhanced. Since magnetic fields cross the shocks vertically, the heat released at the multiple termination shocks is transported along field lines to the upstream of the shocks by heat conduction. The importance of these findings will be discussed in Section~\ref{sec:discussion}.

We summarize a global picture of simulated flare loops in Figure~\ref{fig:flare_illust}. This figure is based on the results of our previous paper and will correspond to an update of the picture based on the standard flare model and \citet{yokoyama1998b}. A noticeable point is that the shocks are formed at different places and affect the density structure. Important features in this study are found in the above-the-loop-top region: oblique fast shocks (multiple termination shocks), backflow of the reconnection outflow, and ``magnetic tuning fork," where the magnetic tuning fork denotes a pair of the sharply bent magnetic field structures in the above-the-loop-top region and will be mentioned later in detail.

\subsection{Above-the-loop-top Oscillation}
Looking at the temporal evolution, we found that the distance between the two arms of the magnetic tuning fork changes quasi-periodically. The oscillation is displayed in Figure~\ref{fig:fork_beta}. The left panels show snapshots of the plasma $\beta$ distribution of the above-the-loop-top region. The right panels show time-sequenced images of plasma $\beta$ and normalized running difference of the total pressure $\Delta p_{\rm tot}/p_{\rm tot}$ obtained along the slit shown in the left panels. The slit is positioned so that its $y$-coordinate is 68~km below the interaction point of the two oblique fast shocks. It is shown that the two arms of the magnetic tuning fork, shown as the two narrow high-$\beta$ regions at the left and right edges, are oscillating with a period of $\sim40$~s (top and bottom rows show the timings when the two arms are closed and open, respectively). This oscillation is hereafter called the ``above-the-loop-top oscillation."

Figure~\ref{fig:fork_beta} also shows that outward-propagating fast waves are quasi-periodically excited when the outward motion of the arms of the magnetic tuning fork terminates (see the time-sequence images of $\Delta p_{\rm tot}/p_{\rm tot}$). These fast waves are what we have already shown in Figure~\ref{fig:sim_intro}. A notable point is that the wave source is localized in the above-the-loop-top region and is very small compared to the system size (less than 10\% of the system size in this simulation). 

A snapshot of the simulation and a schematic illustration of the above-the-loop-top oscillation are shown in Figure~\ref{fig:fork_illust}. When the reconnection outflow impacts on the strong magnetic field region, the flow pattern changes, resulting in backflow (Figure~\ref{fig:fork_illust}(a). See also Figure~\ref{fig:flare_illust}). The backflow (more exactly, the gradient of the dynamic pressure by backflow) pushes the arms of the magnetic tuning fork outward and compresses the magnetic field of the arms. This leads to the generation of outward-propagating fast waves (Figure~\ref{fig:fork_illust}(b). See also the time-sequenced images in Figure~\ref{fig:fork_beta}). Once the magnetic field there becomes strong enough to overcome the backflow, the arms start to move inward, generating inward-propagating fast waves. Although the inward-propagating waves decelerate the backflow, the speed of the backflow quickly recovers, because the speed of the reconnection outflow is almost constant with time. Thus, the same process repeats and the oscillation is maintained. We note that the generation process of fast waves by the backflow-driven magnetic tuning fork is similar to the generation process of sound waves by an externally driven tuning fork. The ``magnetic tuning fork" is named so after its similarity to such a tuning fork.

The oscillation stops when a horizontal fast shock appears in between the two oblique fast shocks (see Figure~\ref{fig:oblique_shock}). The timing of the appearance is also indicated in Figure~\ref{fig:fork_beta}. The horizontal shock more significantly decelerates the reconnection outflow than oblique shocks. Therefore, the backflow of the reconnection outflow, which is essential to maintain the oscillation, becomes slow after the formation of the horizontal shock, leading to the disappearance of the oscillation.

The above-the-loop-top oscillation causes the oscillation of the oblique fast shocks. The temporal evolution of the ratio of the pressures ahead ($p_a$) and behind ($p_b$) one of the oblique shocks is shown in the left panel of Figure~\ref{fig:qpp_qfp} (solid) as an indicator of the shock strength. It can be seen that the shock strength is oscillating with a period of $\sim 40$~s, identical to the period of the above-the-loop-top oscillation. The maximum of the horizontal component of the backflow is also shown (dashed). The quasi-periodic deceleration of the backflow is caused by the inward-propagating fast waves, which are excited by the inward motion of the arms of the magnetic tuning fork (see Figure~\ref{fig:fork_illust}(b)). The right panel displays the wavelet analysis of the coronal fast waves. The normalized running difference of the density $\Delta \rho/\rho$ at the position $(x,y)=(1.5\times10^4~{\rm km},3.9\times10^4~{\rm km})$ (outside the flare loop) is used. A strong power is found at a period of $\sim40$~s, very similar to the period of the above-the-loop-top oscillation. This indicates that QPFs are generated by the above-the-loop-top oscillation. We also note that the shock reflection occurs when the inward-propagating fast waves compress the U-shaped magnetic fields between the arms of the magnetic tuning fork. 

\subsection{Dependence on Magnetic Field Strength}
The dependence of the oscillation period $P$ on the initial plasma $\beta$ (equivalently, the magnetic field strength $B$ in this study) is investigated. Since the oscillation is controlled by the backflow in the above-the-loop-top region, we first look at the dependence of the backflow speed and the size of the above-the-loop-top region.

Figure~\ref{fig:beta_dependence}(a) compares the case with $\beta$=0.06 (Left, strong magnetic field) and the case with $\beta$=0.2 (Right, weak magnetic field), which indicates that in the stronger field case the backflow is faster and the size of the above-the-loop-top region is smaller. The size is defined as the difference in height between the position where the downward flow stops and the top of the magnetic tuning fork (the distance is indicated by the white lines). Figure~\ref{fig:beta_dependence}(b) and (c) display the dependence of the backflow speed $v_{\rm bf}$ and the size $w$, respectively. As an indicator of the backflow speed, we used the time-averaged maximum of the horizontal component of the backflow velocity. The time-averaging is performed during the 72~s after the formation of the oblique fast shocks. The size $w$ is measured at the timing of the second opening of the arms of the magnetic tuning fork. The power law index of $v_{\rm bf}$ is $-0.53$. The power law index of $w$ is $\sim0.44$ (the point for the case of $\beta=0.06$ is not included for this estimation). Figure~\ref{fig:beta_dependence}(d) shows that the dependence of the period $P$. The power law slope is $\sim 0.96$. This power law slope is very similar to that of the timescale determined by the backflow: $w/v_{\rm bf}\propto \beta^{0.44+0.53}=\beta^{0.97}$, which indicates that the oscillation period approximately scales as the timescale determined by the backflow.

As shown in Figure~\ref{fig:beta_dependence}(b), the backflow speed is of the order of the Alfv\'en speed (the dashed line indicates $0.45V_{A,0}\propto \beta^{-0.5}$, where $V_{A,0}$ is the initial Alfv\'en speed in the corona). The reconnection outflow speed $v_{\rm outflow}$ is almost the same as $V_{A,0}$. The reason why the backflow behind the multiple termination shocks is Alfv\'enic is that the deceleration of the reconnection outflow by oblique shocks is inefficient. This Alfv\'enic backflow drives the above-the-loop-top oscillation.

The dependence of the size will be explained as follows. Assuming the conservation of mass and the quasi-steady state, the mass flux of the reconnection outflow and the mass flux horizontally carried by the backflow will be balanced:
\begin{align}
 \rho_{1}v_{\rm outflow}d \sim \rho_{2}v_{\rm bf}w,
\end{align}
where $\rho_1$ and $\rho_2$ are the average densities in the reconnection outflow and in the above-the-loop-top region, respectively, and $d$ is the width of the reconnection outflow. The opening angle of the reconnection outflow $\theta\sim d/L$ is almost equivalent to the reconnection rate \citep{petschek1964}. In all the simulations the normalized reconnection rate is approximately 0.06. This gives $d\sim0.06L$. The termination shocks are isothermal shocks owing to the heat conduction. For this reason, the compressional ratio through the shocks roughly becomes $\rho_2/\rho_1\sim M^2$, where $M=v_{\rm outflow}/C_s$ is the acoustic Mach number and $C_s$ is the sound speed in the outflow. Here we neglected the effects of the shock angle and assumed that the plasma beta in the reconnection outflow $\beta_{\rm outflow}$ is much larger than unity (the condition $\beta_{\rm outflow}\gg1$ is valid in our simulations). If we consider the effect of the heat conduction, the scaling of the Mach number will be $M\propto \beta^{-2/7}L^{-1/7}$ (see Appendix, Relation~\ref{eq:mach}). From this, we get $\rho_2/\rho_1\sim \beta^{-4/7}L^{-2/7}$. Finally, we obtain
\begin{align}
w\sim \frac{\rho_1}{\rho_2} \frac{v_{\rm outflow}}{v_{\rm bf}} d \sim 2\frac{\rho_1}{\rho_2}d\propto \beta^{4/7}L^{9/7}=\beta^{0.57}L^{1.3}.
\end{align}
The predicted slope is consistent with the numerical results (Figure~\ref{fig:beta_dependence}(c)). This scaling relation will be invalid when $d\simeq w$. It is shown in Figure~\ref{fig:beta_dependence}(c) that the width of the outflow $d$ in the case of $\beta=0.06$ is similar to the size $w$, which explains the deviation of the point.

From the theoretical estimation above, the timescale of the backflow (and therefore the period $P$) is expected to scale as 
\begin{align}
P\propto \frac{w}{v_{\rm bf}}\propto \beta^{15/14}L^{9/7} = \beta^{1.1}L^{1.3}
\end{align}
(we only look at the $\beta$ dependence). Figure~\ref{fig:beta_dependence}(d) shows that the theoretical scaling of $P$ is consistent with our simulations.

\section{Discussion}\label{sec:discussion}
% summary
We carried out MHD simulations of a solar flare in which essential physics for solar flares such as magnetic reconnection, heat conduction, and chromospheric evaporation were included. Our model revealed that flare loops and the above-the-loop-top region are full of shocks and waves. From our simulations, we discovered the local oscillation of the above-the-loop-top region (above-the-loop-top oscillation) and the generation of quasi-periodic propagating fast mode magnetoacoustic waves (QPFs) from such oscillating regions. It was found that the above-the-loop-top oscillation is controlled by the backflow of the reconnection outflow in the above-the-loop-top region. This means that the wave source is localized and very small compared to the flare loop size (less than 10\% of the flare loop size). It was revealed that the termination shock structure has a significant impact on the maintenance and termination of the oscillation. The generation process of QPFs is found to be similar to the sound wave generation by an externally driven tuning fork.

Many previous models for QPFs do not clearly specify the physical origins of exciters of coronal waves and have been used for the investigation of the propagation and dispersive nature of waves of interest \citep[e.g.][]{ofman2011,pascoe2013}. Using MHD simulations, we revealed that the reconnection outflow (more exactly, the backflow of the outflow) can act as an exciter of coronal waves. We also showed that waves can be spontaneously generated even with a quasi-steady reconnection outflow. It may be possible that a time-dependent and oscillatory reconnection process also leads to the generation of QPF s\citep{kliem2000,mclaughlin2009,murray2009}. It is claimed on the basis of two-dimensional MHD simulations that the quasi-periodic ejection of plasmoids could also lead to QPFs \citep{yokoyama1998a,yang2015}. However, the behavior of reconnection in three-dimensions remains poorly understood. The initial magnetic field configuration may also affect the oscillation processes. We will investigate the influence of the three-dimensional reconnection in a more realistic magnetic configuration in our future papers.

Previous studies of flare loop oscillations mainly focus on the standing (M)HD waves in the loops \citep[see a review by][]{nakariakov2009}. In this study, however, we discovered a cyclic process that is controlled by the flow confined in the above-the-loop-top region, not by any standing waves. This finding has a significant impact on the interpretation of oscillations. For instance, if we consider that oscillation is caused by a standing acoustic wave \citep{nakariakov2004}, the period can be interpreted as
\begin{align}
P\propto \frac{L}{C_s}\propto LT^{-1/2} \propto B^{-3/7}L^{6/7}\propto B^{-0.43},
\end{align}
where $C_s\propto T^{1/2}$ is the sound speed, and we used the scaling law of the flare temperature ($T\propto B^{6/7}L^{2/7}$) by \citet{yokoyama1998b}. Here we only focus on the dependence on the magnetic field strength for clarity. However, in the case of the above-the-loop-top oscillation, the period can be written as
\begin{align}
P\propto \frac{w}{v_{\rm bf}}\propto \beta^{15/14} \propto B^{-2.1},
\end{align}
which gives a different scaling. Hence, it is crucial to correctly identify the oscillation mechanism before one derives the physical parameters from observations. The development of methods to distinguish various kinds of oscillations including the above-the-loop-top oscillation is required. This will greatly advance the solar and stellar coronal seismology. We infer that a large dispersion of the observed oscillation period (a few 10~sec to a few 100~sec) partly reflects the strong dependence on the magnetic field strength.

\citet{shibata2002} developed a theory to estimate the physical parameters of solar and stellar flares from the observable parameters (the emission measure and the temperature derived from soft X-ray observations), on the basis of the results of MHD simulations of a solar flare by \citet{yokoyama2001}. Their theory is based on a reconnection model for flares and is derived under the assumption that the pressure in the flare loops is balanced with the magnetic pressure outside. We derived a scaling relation of the oscillation period on the basis of similar MHD simulations. Our scaling relation and their theory are both based on the reconnection physics but describe different aspects of flares. Therefore, a combination of both our scaling relation and their theory will provide a powerful reconnection-based method to diagnose the solar and stellar coronae.

%relation to QPP in the nonthermal emissions
We found that the termination shocks are also quasi-periodically oscillating because of the above-the-loop-top oscillation (Figure~\ref{fig:qpp_qfp}). It has been argued that termination shocks could be a promising site for particle acceleration \citep{tsuneta1998,nishizuka2013,chen2015}, and could be related to the above-the-loop-top hard X-ray source \citep{masuda1994,krucker2010,oka2015}, although the detailed acceleration process in such low Mach number and high-$\beta$ shocks should be studied in more detail (for recent studies about the electron acceleration, see \citet{matsukiyo2011,guo2014}) If this is true, the quasi-periodic oscillation of the multiple termination shocks found in this study could lead to QPPs in the nonthermal emissions through the quasi-periodic variation of the efficiency of particle acceleration. QPPs in the nonthermal emissions have been commonly observed during flares \citep{aschwanden2002,nakariakov2009}, but the origin has been puzzling. Our study could provide a possible solution for this. In addition, the simulations showed that the oscillation of the multiple termination shocks and QPFs can have a common origin. On this basis, we suggest a new picture in which QPFs and QPPs in the nonthermal emissions have a common origin. Thermal emissions may also respond to the variation of the efficiency of acceleration through thermalization of nonthermal particles, showing QPPs.

\begin{figure}
\epsscale{0.85}
%\plotone{te_ro_line_beta008_combine02.eps}
\plotone{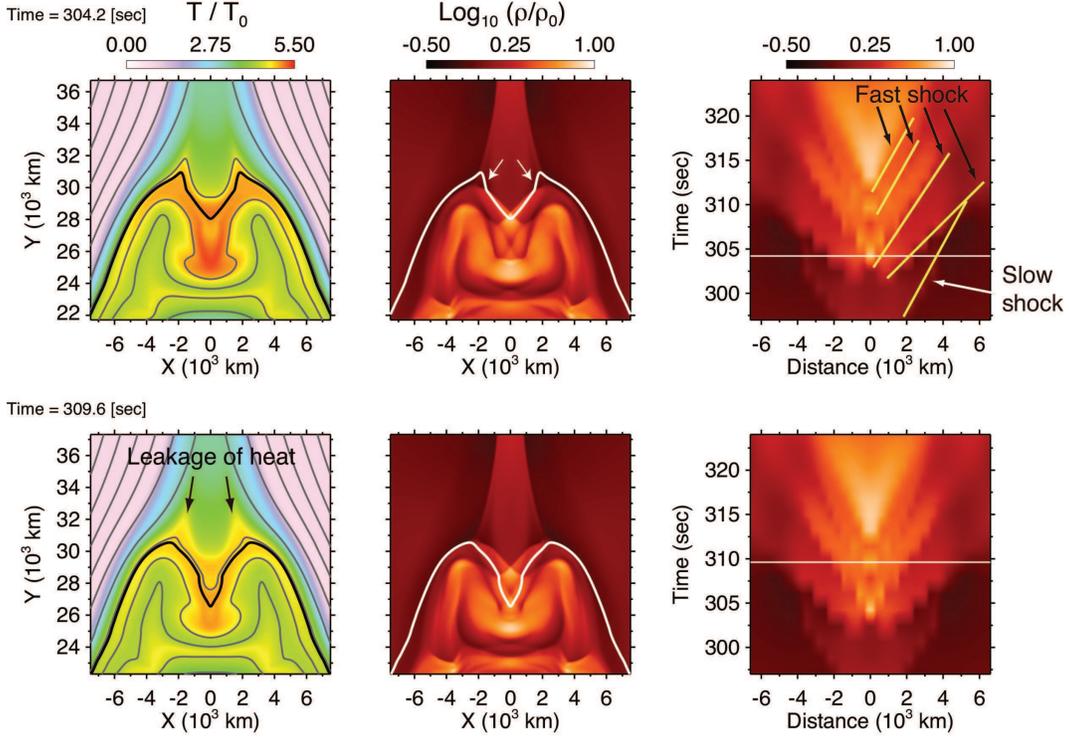}
\caption{Passage of a specific field line through the multiple termination shocks. Left panels: Snapshots of the temperature and density distributions of the above-the-loop-top region. The black field lines in the temperature map and the white field lines in the density map denote the same field line. Right panels: Time-sequenced images obtained along the tracked field line (density). The white solid lines denote the timings of the snapshots in the left panels. The shocks which the field line crosses in the region $x>0$ are denoted by the yellow solid lines. The segments between a slow shock and a fast shock are indicated by arrows in the density map. \label{fig:te_ro_line}}
\end{figure}

\begin{figure}
\epsscale{0.9}
\plotone{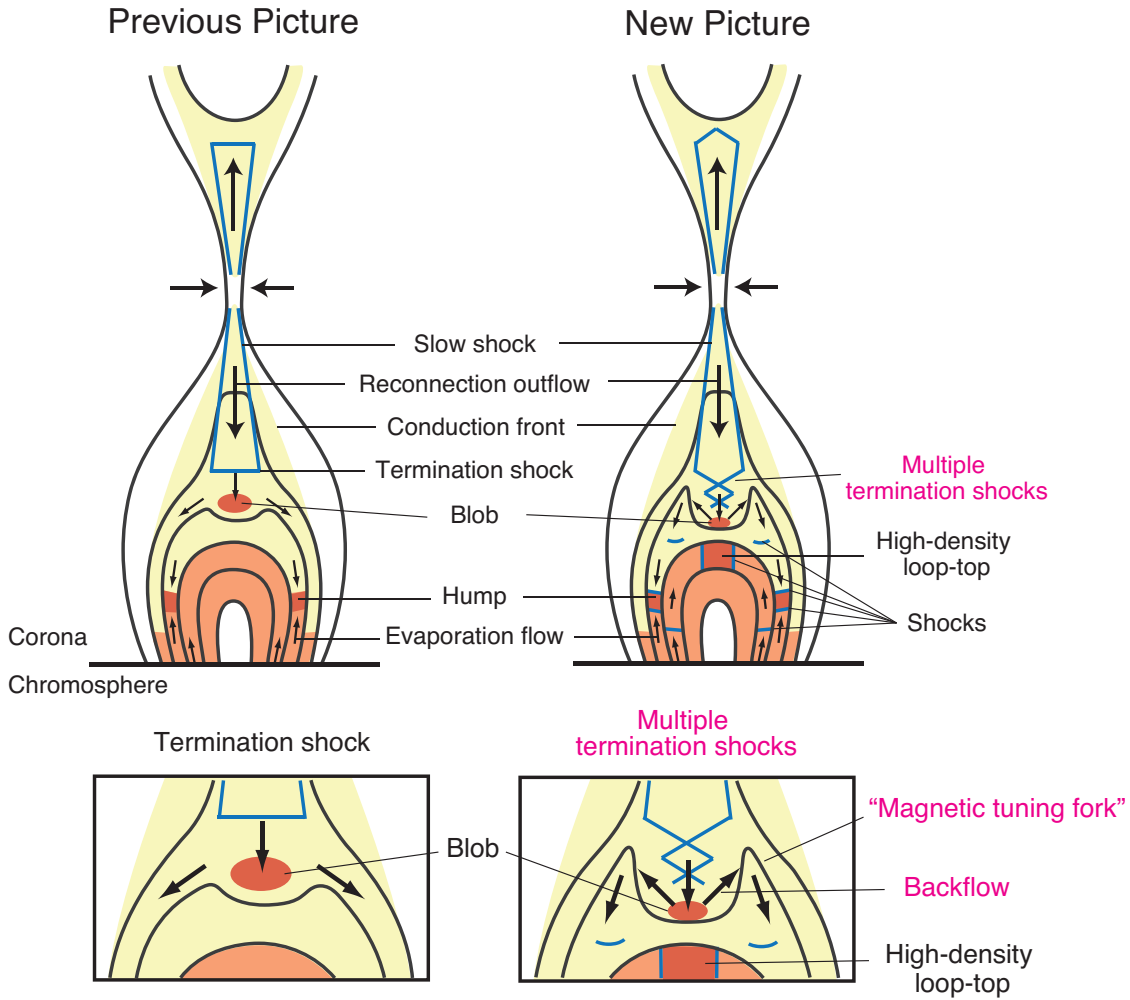}
%\plotone{test.eps}
\caption{Comparison of a previous picture of flare loops based on a standard flare model and \citet{yokoyama1998b} model (Left) with a new picture based on our simulations (Right). Black solid lines denote magnetic field lines. Pale orange regions denote evaporated plasma. Orange regions indicate dense regions. Pale yellow regions denote regions inside the conduction fronts. Blue solid lines indicate shocks. Flows are denoted by arrows. A detailed description of the new picture is given in \citet{takasao2015}. Enlarged images of the above-the-loop-top regions are displayed at the bottom.\label{fig:flare_illust}}
\end{figure}

\begin{figure}
\epsscale{0.9}
%\plotone{slice_looptop_beta008_01.eps}
\plotone{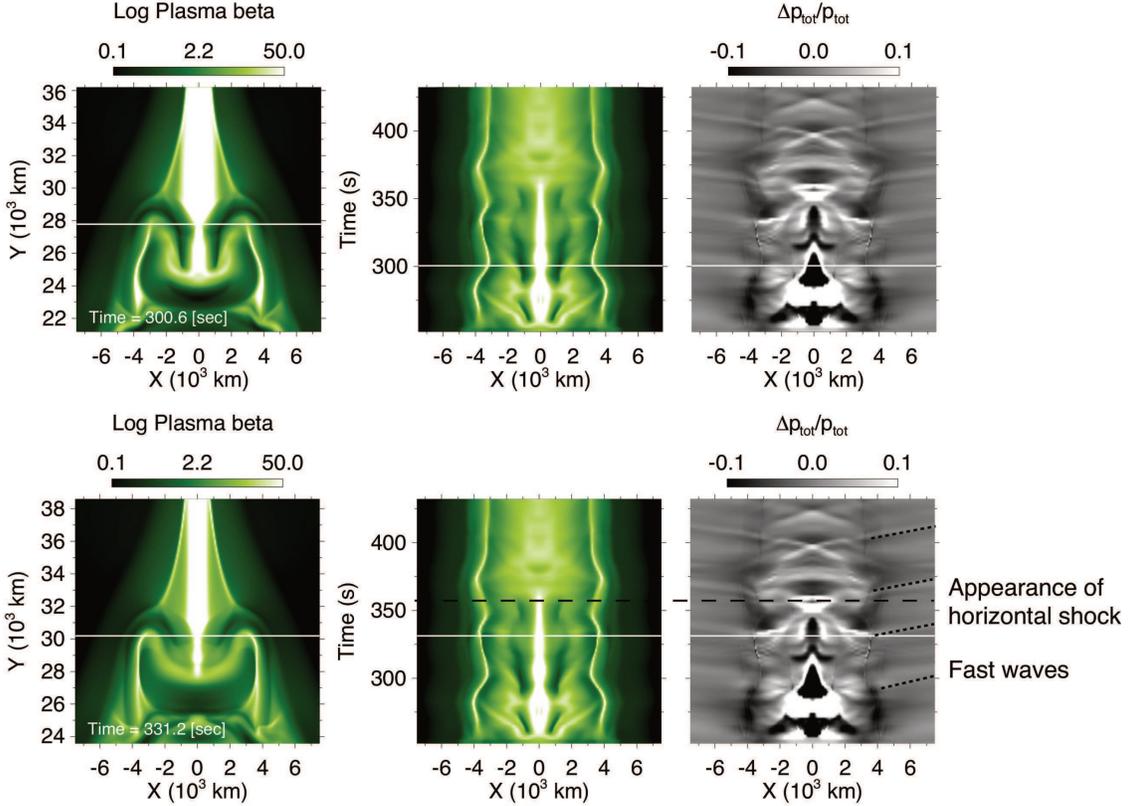}
%\plotone{test.eps}
\caption{Above-the-loop-top oscillation (the initial $\beta$ is 0.08). Left: Snapshots of the distribution of $\log_{10}\beta$ of the above-the-loop-top region. Right: Time-sequenced images of $\log_{10}\beta$ and $\Delta p_{tot}/p_{tot}$ obtained along the slit shown in the left panels. The slit used is positioned so that its $y$-coordinates is 68~km below the interaction point of the two oblique shocks. The horizontal lines in the time-sequenced images denote the timings of the snapshots in the left panels. An animation of this figure is also available in the online journal. \label{fig:fork_beta}}
\end{figure}

\begin{figure}
\epsscale{0.8}
\plotone{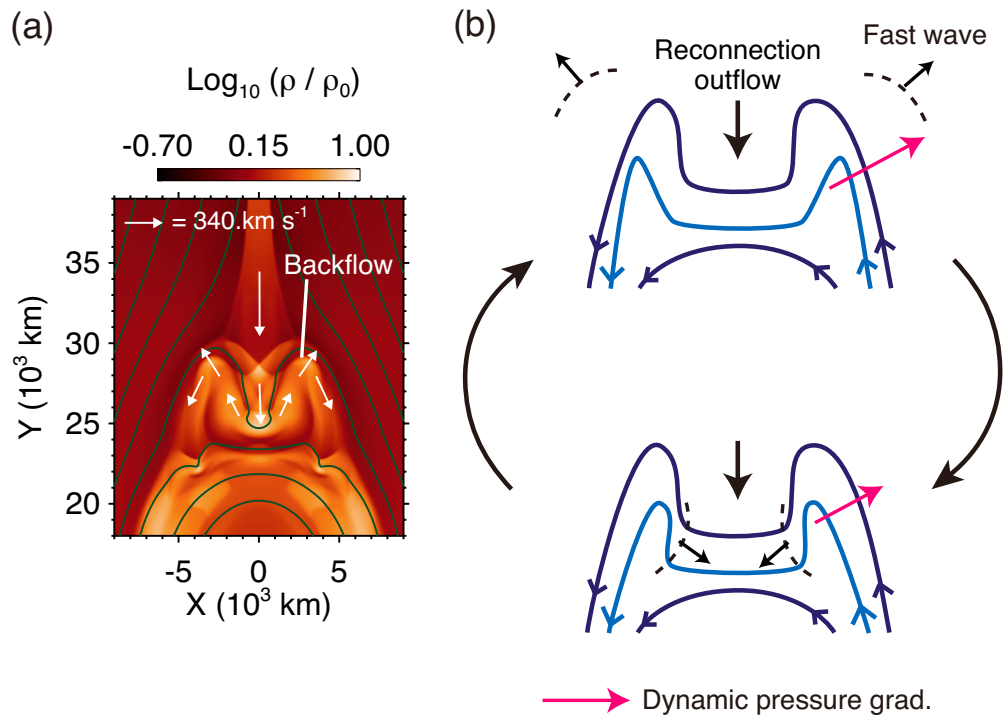}
%\plotone{test.eps}
\caption{(a) Backflow of the reconnection outflow in the above-the-loop-top region. (b) Schematic illustration of the above-the-loop-top oscillation. \label{fig:fork_illust}}
\end{figure}

\begin{figure}
\epsscale{0.9}
\plotone{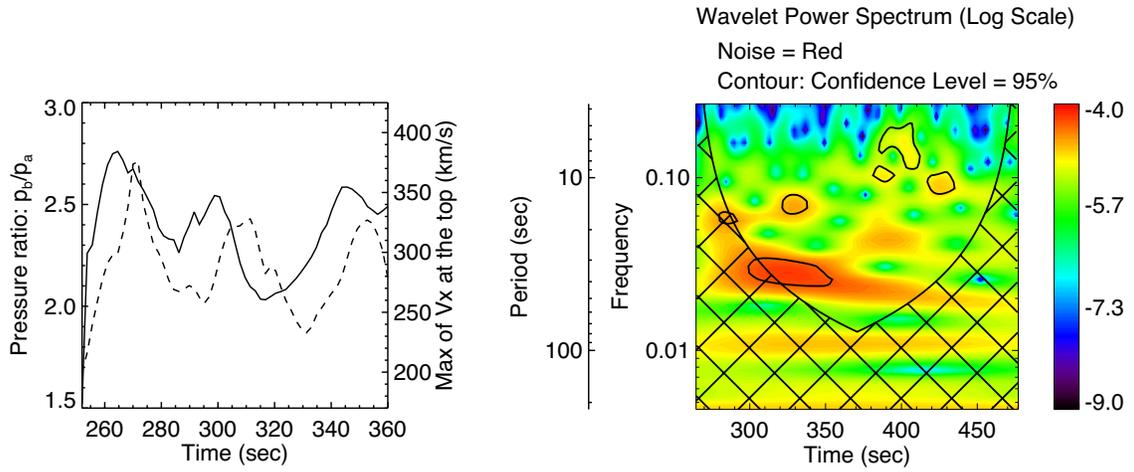}
%\plotone{test.eps}
\caption{Relation between the above-the-loop-top oscillation and coronal waves (a) Temporal evolution of the ratio of the pressures ahead ($p_a$) and behind ($p_b$) a topmost oblique shock (solid) and the maximum of the horizontal component of the backflow (dashed). (b) Wavelet analysis of the coronal fast waves. The normalized running difference of the density at the position $(x,y)=(1.5\times10^4~{\rm km}, 3.9\times10^4~{\rm km})$ is used. \label{fig:qpp_qfp}}
\end{figure}

\begin{figure}
\epsscale{0.9}
\plotone{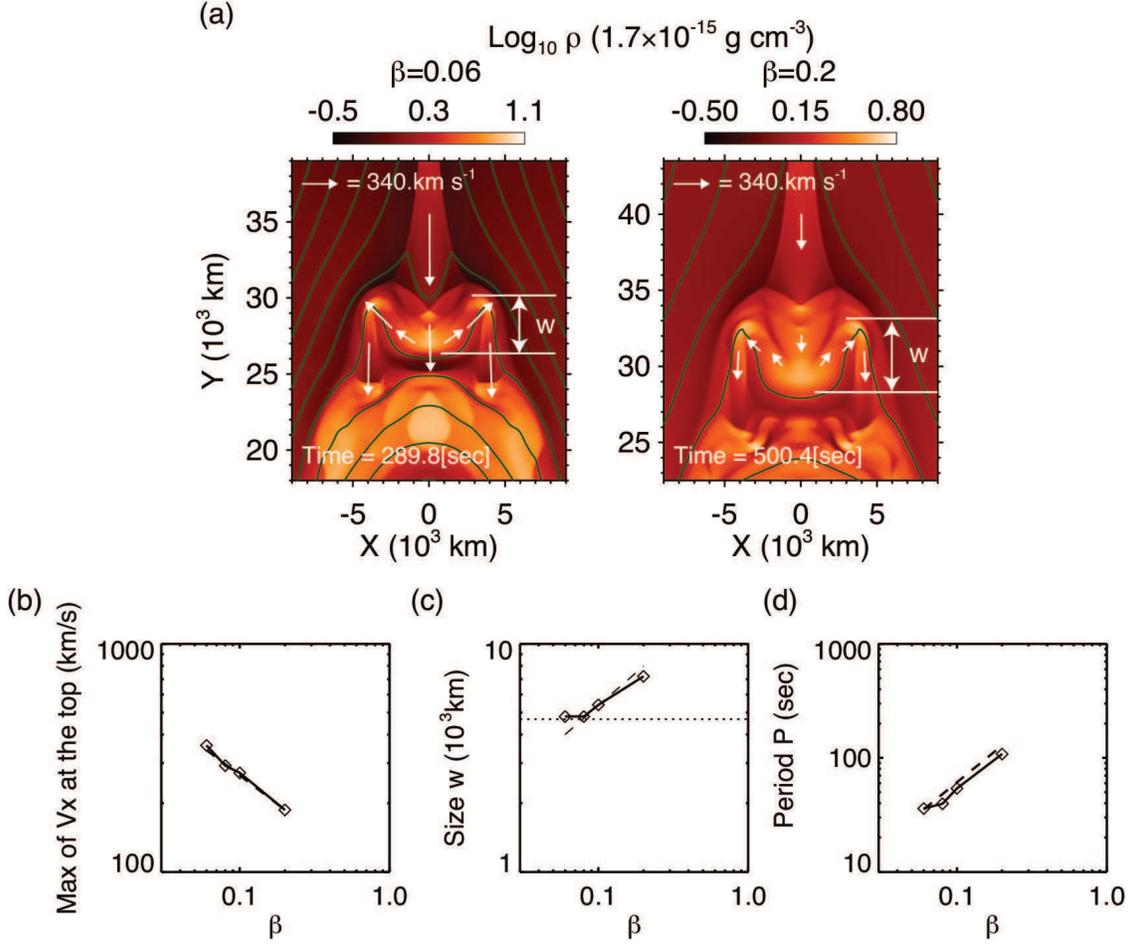}
\caption{(a) Comparison of the cases with $\beta=0.06$ (Left) and $\beta=0.2$ (Right). (b-d) Plasma $\beta$ dependence of the maximum of the horizontal component of the backflow ($v_{\rm bf}$) (b), the size of the above-the-loop-top region ($w$) (c), and the period ($P$) (d). The period is estimated from the time interval between the initial and second peaks seen in the pressure ratio $p_b/p_a$. In the panel~(b), the dashed line denotes $0.45V_{A0}(\propto \beta^{-0.5})$. In the panels~(c) and (d), the dashed lines indicate the slope of $\beta^{4/7}$ and $\beta^{15/14}$, respectively. The horizontal dotted line in the panel~(c) denotes the width of the reconnection outflow. The size of the above-the-loop-top region is defined as the difference in height between the position where the downward flow stops and the top of the magnetic tuning fork (the distance is indicated by the white lines in the panel~(a)). The size $w$ is measured at the timing of the second open of the arms of the magnetic tuning fork. \label{fig:beta_dependence}}
\end{figure}

%preheating
We showed that a field line can simultaneously cross multiple shocks when it passes through the multiple termination shocks  (Figure~\ref{fig:te_ro_line}). This indicates that one can expect more chances for particle acceleration at the shocks than the case with a single horizontal termination shock, which is commonly assumed in the standard flare model. The segment between a slow shock and a topmost fast shock (indicated in the density map of Figure~\ref{fig:te_ro_line}) could be a good site for the Fermi acceleration, because these shocks are approaching each other. It is also found that the temperature in the upstream of the topmost oblique fast shocks (including the segments between a slow shock and a fast shock) is enhanced by heat conduction (indicated in the temperature map of Figure~\ref{fig:te_ro_line}). For an efficient particle acceleration, the preheating of plasma before the acceleration at termination shocks may be necessary. \citet{tsuneta1998} considered that the heating by the slow shocks attached to the reconnection outflow will provide a method for preheating. We consider that the leakage of the heat released at the multiple termination shocks will also contribute to the preheating.

\acknowledgements
S.T. acknowledges support by the Research Fellowship of the Japan Society for the Promotion of Science (JSPS). This work was supported by a Grant-in-Aid from the Ministry of Education, Culture, Sports, Science and Technology of Japan (No. 25287039). We are grateful to SDO/AIA team for providing the data used in this study. Wavelet software was provided by C. Torrence and G. Compo, and is available at URL: \url{http://atoc.colorado.edu/research/wavelets/}.

{\it Facillities:} \facility{SDO}

\appendix
\section{Scaling of the Fast-mode Mach Number of Reconnection Outflow Jet}\label{sec:appendix}
The scaling of the fast-mode Mach number of the reconnection outflow jet ($M_{\rm FM}$) will be briefly discussed. $M_{\rm FM}$ can be written as
\begin{align}
M_{\rm FM}=\frac{v_{\rm outflow}}{c_{f,out}}\simeq \frac{v_{A,in}}{c_{s,out}}\propto \beta_{in}^{-1/2}\left( \frac{T_{out}}{T_{in}}\right)^{-1/2}
\end{align}
where $v_{\rm outflow}$ indicates the outflow speed. The quantities in the inflow and outflow regions are denoted with the subscripts ``in" and ``out," respectively. Here we assume that the reconnection outflow is high-$\beta$ and the fast-mode wave speed $c_{f,out}$ and sound speed $c_{s,out}$ are similar. If we neglect the effect of heat conduction, the temperature increase at the slow shock $\Delta T$ can be estimated as
\begin{align}
\frac{\Delta T}{T_{in}} \simeq \beta_{in}^{-1}\gg 1 \hspace{3mm} 
\end{align}
for low-$\beta$ plasma, so $T_{out}/T_{in}\simeq \beta_{in}^{-1}$. However, with the effect of heat conduction, 
\begin{align}
T_{out} \propto \beta_{in}^{-3/7}n_{in}^{2/7}L^{2/7}
\end{align}
\citep[see][]{yokoyama2001,shibata2002}. The difference in scaling of the temperature leads to the different scalings of the fast-mode Mach number of the outflow jet:
\begin{eqnarray}
M_{\rm FM}\propto \left\{ \begin{array}{ll}
{\rm const.} & \textrm{(without heat conduction)} \\
\beta_{in}^{-2/7}n_{in}^{-1/7}L^{-1/7}\propto B^{4/7}n_{in}^{-3/7}L^{-1/7} & \textrm{ (with heat conduction)} \\
\end{array} \right.\label{eq:mach}
\end{eqnarray}
This means that heat conduction has a significant effect in increasing the fast-mode Mach number. This effect is also pointed out by \citet{seaton2009}. 

\begin{figure}
\epsscale{0.8}
\plotone{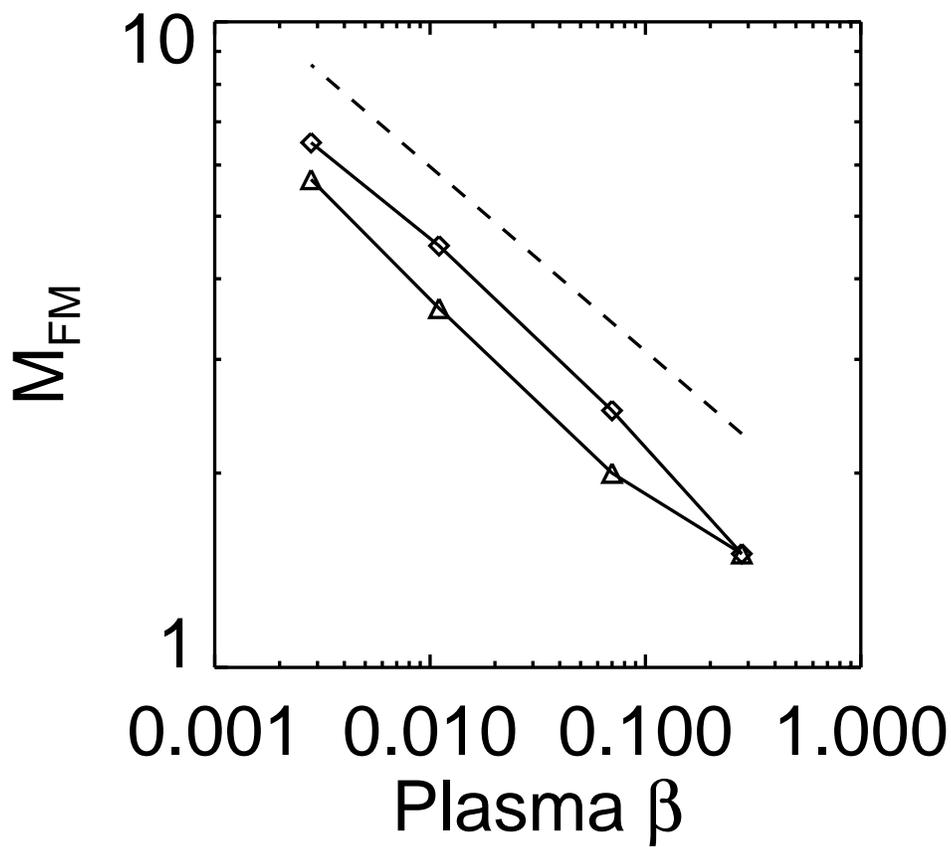}
\caption{Dependence of the fast-mode Mach number of the reconnection outflow ($M_{\rm FM}$) on the plasma $\beta$. The data points indicated by diamonds and triangles are from \citet{seaton2009}. Diamonds and triangles denote the data points for the normalized conduction coefficient $\lambda^*=$0.03 and 0.003, respectively (see the text for the definition of $\lambda^*$). The dashed line indicates the slope $\propto \beta_{in}^{-2/7}$.  \label{fig:seaton}}
\end{figure}

The dependence of $M_{\rm FM}$ on the plasma $\beta$ is analytically investigated by \citet{seaton2009}. They define the normalized thermal conduction coefficient $\lambda^*$ as the ratio of the energy loss due to heat conduction ($F_{\rm cond}$) to the energy input by Poynting flux carried at the Alfv\'en speed $V_{\it A}$ into the current sheet ($F_{\rm P,Alfven}$): $\lambda^* \equiv F_{\rm cond}/F_{\rm P,Alfven}$. \citet{yokoyama1998b,yokoyama2001} found that the assumption that the heating by Poynting flux carried at the reconnection inflow speed $v_{\rm inflow}$ balances with the conduction cooling provides a good approximation. Considering this, we obtained the following restriction on $\lambda^*$:
\begin{align}
\lambda^* = \frac{F_{\rm cond}}{F_{\rm P,Alfven}} =  \frac{F_{\rm cond}}{F_{\rm P,inflow}}  \frac{F_{\rm P,inflow}}{F_{\rm P,Alfven}} \simeq O(1) \frac{v_{\rm inflow}}{V_{\it A}} \simeq 0.001-0.1,
\end{align}
where we assume that the nondimensional reconnection rate $v_{\rm inflow}/V_{\it A}$ is in the range of approximately 0.001--0.1, considering observations of solar flares \citep{narukage2006,takasao2012}. Figure~\ref{fig:seaton} displays the dependence of $M_{\rm FM}$ on the plasma $\beta$ from \citet{seaton2009}. The data points denoted by triangles and diamonds are for the cases with $\lambda^*$=0.003 and 0.03, respectively, so $0.001<\lambda^*<0.1$. The dashed line indicates the slope of the scaling relation~(\ref{eq:mach}), proportional to $\beta_{in}^{-2/7}$. The figure shows that the scaling relation is consistent with their analytical results. Therefore, it is shown that the simple argument here can explain the dependence of $M_{\rm FM}$.

\end{document}